\documentclass[conference]{IEEEtran}

\usepackage{graphicx}
\usepackage{algorithm} 
\usepackage{algorithmic}
\usepackage{cite}
\usepackage{fixltx2e}

\begin{document}

\title{Pattern Tree-based XOLAP Rollup Operator\\for XML Complex Hierarchies}

\author{\IEEEauthorblockN{Marouane Hachicha}
\IEEEauthorblockA{Universit\'{e} de Lyon (ERIC Lyon 2)\\
5 avenue Pierre Mend\`{e}s-France\\
69676 Bron Cedex, France\\
Email: marouane.hachicha@univ-lyon2.fr}
\and
\IEEEauthorblockN{J\'{e}r\^{o}me Darmont}
\IEEEauthorblockA{Universit\'{e} de Lyon (ERIC Lyon 2)\\
5 avenue Pierre Mend\`{e}s-France\\
69676 Bron Cedex, France\\
Email: jerome.darmont@univ-lyon2.fr}
}

\maketitle

\begin{abstract}
With the rise of XML as a standard for representing business data, XML data warehousing appears as a suitable solution for decision-support applications. In this context, it is necessary to allow OLAP analyses on XML data cubes. Thus, XQuery extensions are needed. To define a formal framework and allow much-needed performance optimizations on analytical queries expressed in XQuery, defining an algebra is desirable. However, XML-OLAP (XOLAP) algebras from the literature still largely rely on the relational model. Hence, we propose in this paper a rollup operator based on a pattern tree in order to handle multidimensional XML data expressed within complex hierarchies.
\end{abstract}

\section{Introduction}

In many institutions, decision-support applications require external data. In this context, the Web is a tremendous data source and Web farming \cite{Hackathorn99} is more and more casual. As a consequence, a new trend toward on-line data warehousing is currently emerging, including approaches such as XML warehousing \cite{ZhangWLZ05}.

The XML language is indeed becoming a standard for representing business data \cite{BeyerCCOPX05}. Moreover, it is particularly adapted for modeling so-called complex data \cite{DarmontBRA05} from heterogeneous sources, and particularly the Web. Thus, many studies aim at extending the XQuery language \cite{BoagCFFRS07} with OLAP (On-Line Analytical Processing)-like queries (grouping, aggregation, etc.) \cite{BeyerCCOPX05,BorkarC04,Kay06}. Such extensions should not only allow classical OLAP analyses, but also take the specificities of XML into account, e.g., ragged hierarchies \cite{BeyerCCOPX05} that would be intricate to handle in a relational environment.

In this context, we are working to propose an OLAP~algebra over multidimensional XML data~(XML data modeled in multidimensional way). On the long run, we are actually aiming at three objectives: contribute to define a formal framework that does not currently exist in the XOLAP~\cite{WangLHG05} context; support the effort for extending the XQuery language to allow OLAP queries, especially with XML-specific operators; allow query optimization for OLAP XQueries. Native-XML DBMSs (Database Management Systems), though in constant progress, are indeed limited in term of performance and would greatly benefit from automatic query optimization, especially for costly analytical queries.

In this paper, we particularly focus on the first objective. In a previous work, we have expressed classical (structural, set and granularity-related) OLAP operators over multidimensional XML data organized in simple hierarchies~\cite{datax08hmd} with the TAX XML algebra~\cite{JagadishLST01}. The next step is now to take XML specifics into account and propose operators for data organized in ragged, complex hierarchies. TAX, as many other XML algebras, is based on pattern trees~\cite{AmerYahiaCLS01} to model user queries. Further combining TAX operators to process hierarchies with unpredictable structures would require to handle combinations of many pattern trees. On the other hand, a more straightforward way to achieve our goal is to directly work at the pattern tree level and design a single, ad-hoc pattern tree. Hence, we propose a pattern tree model with advanced matching capabilities, including aggregation, grouping and ordering; and illustrate its use through a rollup operator that applies onto complex hierarchies. This marks a first step in defining a full set of pattern tree-based XOLAP operators.

The remainder of this paper is organized as follows. In Section~\ref{background}, we formally define pattern trees and related concepts. In Section~\ref{Pattern-xml-algebras}, we survey the pattern trees that are used in XML algebras. In Section~\ref{complex-hierarchies}, we formally define the complex hierarchies we want to handle. In Section~\ref{A-Pattern-Tree-for-XOLAP}, we introduce our pattern tree-based rollup operator. We finally conclude this paper and discuss research perspectives in Section~\ref{Conclusion}.

\section{Background}
\label{background}

We define in this section the main concepts that lie behind XML algebras, i.e., XML data trees and subtrees, pattern trees and the operations of matching and embedding. 

\subsection{XML Data Trees and Subtrees} 
\label{XML-data-tree-and-subtree}

A data tree \textit{t} models an XML document or a document fragment. It may be defined as a triple \textit{t} = $\left(\textit{r}, \textit{N}, \textit{E}\right)$, where \textit{N} is the set of nodes, \textit{r}~$\in$~\textit{N} is the root of \textit{t}, and \textit{E} is the set of edges stitching together couples of nodes (\textit{n}$_{i}$, \textit{n}$_{j}$) $\in$~\textit{N}. 

Given an XML data tree \textit{t}~=~(\textit{r}, \textit{N}, \textit{E}) and $e \in E$ an edge connecting two nodes $(n_i, n_j)$. \textit{t}$^{'}$~=~(\textit{r}$^{'}$, \textit{N}$^{'}$, \textit{E}$^{'}$) is a subtree of \textit{t} \textit{iff} the following conditions are satisfied:~\textit{N}${'}$$\subseteq$~\textit{N}; there exists an edge $e^{'} \in E^{'}$ connecting two nodes $(n_i^{'}, n_j^{'})$ such that $n_i = n_i^{'}$ and $n_j = n_j^{'}$.

\subsection{Pattern Trees}
\label{Pattern-trees}

A pattern tree~\textit{pt}, also called tree pattern or tree pattern query (TPQ) \cite{AmerYahiaCLS01} is a pair $\left(t, F\right)$ where:~(1)~\textit{t} is a tree (\textit{r}, \textit{N}, \textit{E}). An edge may either be a parent-child (\textit{pc} for short, simple~edge~/~in XPath) node relationship or an ancestor-descendant (\textit{ad} for short, double~edge~//~in XPath) node relationship;~(2)~\textit{F} is a formula that specifies constraints on node values. More explicitly, $F$ is a boolean combination of predicates on node values.
		
Basically, a pattern tree captures a useful fragment of XPath \cite{FlescaFM03}. But it can also be seen as the translation of a user query formalized in natural language or in an XML query language such as XQuery \cite{ChenJLP03}. Translating an XML query plan into a pattern tree is not a simple operation. Some XQueries are written with complex combinations of XPath and FLWOR expressions, and imply more than one pattern tree. Such queries must be broken up into several patterns trees. Only a single XPath expression can be translated into a single pattern tree. The more a query is difficult, the more its translation in pattern tree(s) is complex \cite{MichielsMS07}. To this aim, starting from patterns to express user queries in a first stage, and optimizing them in a second stage is a very effective solution for XML query optimization.

\subsection{Matching and Embedding}
\label{Matching-and-embedding}

Answers for pattern trees (named witness trees in TAX) are formalized through one or multiple matchings. Matching a pattern tree \textit{pt} into an XML data tree {t} is a function \texttt{f:pt$\rightarrow$t} that maps nodes of \textit{pt} to nodes of \textit{t} such that:~(1)~structural relationships are preserved, i.e., if nodes $(x, y)$ are related in $t$ through a $pc$ node relationship (respectively an \textit{ad} node relationship), their counterparts $(x^{'}, y^{'})$ in $pt$ must be related through a $pc$ node relationship (respectively an \textit{ad} node relationship) too; and~(2)~formula \textit{F} of \textit{pt} must be satisfied. 

Embedding a pattern tree $pt$ into a data tree $t$ is a function $g: pt \rightarrow t$ that maps each node of $pt$ to nodes of $t$ such that structural relationships (\textit{pc} and \textit{ad}) are preserved. The difference between embedding and matching is that embedding maps a pattern tree against a data tree \emph{structure} only, whereas matching maps a pattern tree against a data tree structure \emph{and contents}~\cite{LakshmananRWZ04}. In the remainder of this paper, we use the more general term matching when referring to mapping pattern trees against data trees.

\subsection{Example}
\label{def-example}

\begin{figure*}[!t]
\centering
\includegraphics[width=11cm]{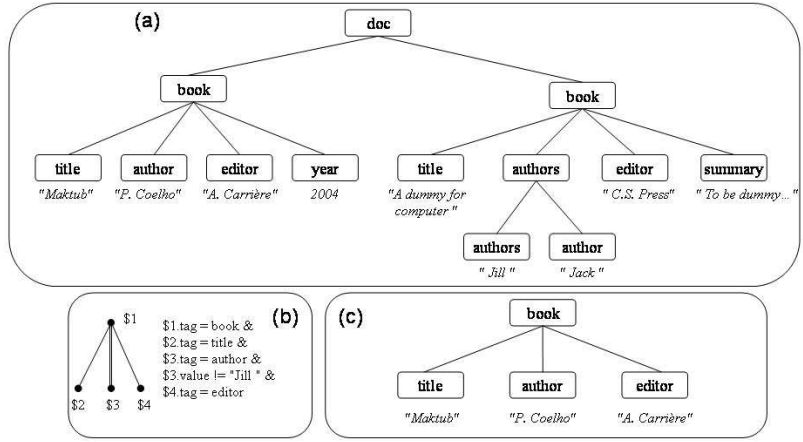}
\caption{XML data (a), pattern (b) and witness trees (c)}
\label{Doc-Books.jpg}
\end{figure*}

For comprehensibility, let us consider the XML data tree from Figure~\ref{Doc-Books.jpg}(a) that represents a collection of books. Root \textit{doc} unites \textit{books} described by their \textit{titles}, \textit{authors}, \textit{editors}, \textit{years} and \textit{summaries}. Data trees nodes are connected by simple edges (/), i.e., \textit{pc} relationships. Books are not necessarily described the same way. For instance, a summary may not be present in all books. Some books can be written by more than one author.

The pattern tree from Figure~\ref{Doc-Books.jpg}(b) selects book titles, authors, and editors. Moreover, formula $F$ indicates that author must be different from Jill. Matching this pattern tree against the data tree from Figure~\ref{Doc-Books.jpg}(a) outputs the data tree (or witness tree) from Figure~\ref{Doc-Books.jpg}(c). Only one book is selected, since the other one ($title = $~``A dummy for a computer'') is written by $author =$ ``Jill'', which contradicts formula $F$.

Finally, the \textit{ad} relationship $\$1//\$3$ in Figure~\ref{Doc-Books.jpg}(b)'s pattern tree is correctly taken into account. The book element ($title =$ ``A dummy for a computer'') is indeed not disqualified because of its structure, but because one of its authors is Jill. It this author was Gill, the book would appear in output.

\section{Pattern Trees Used in XML Tree Algebras}
\label{Pattern-xml-algebras}

The aim of an XML tree algebra is to feature a set of algebraic operators to manipulate and query XML data tree structures. The output of a query formulated over a tree must also be a data tree that respects a tree structure, i.e., a pattern tree.

First XML algebras have appeared in 1999 \cite{beech99} in conjunction with efforts aiming to define a powerful query language for XML \cite{ChamberlinRF-quilt00}. Note that these XML algebras have appeared before the first specification of XQuery, which is regarded as the most popular XML query language, in 2001 \cite{ChamberlinFRSS2001}.

TAX is one of the most popular XML tree algebras~\cite{JagadishLST01}. The TAX pattern tree represents the most basic pattern tree used in an algebraic context. It preserves \textit{pc}-\textit{ad} relationships from the input ordered data tree in output and it satisfies the formula associated to the pattern. The examples from Figure~\ref{Doc-Books.jpg}(a), (b) and (c) correspond to TAX data, pattern and witness trees, respectively.

Providing more matching options for edges connecting output nodes allows a more efficient extraction of these nodes when matching the relevant pattern tree. An important limitation of the TAX pattern tree comes in case of absence of one node in the subtrees matched with the pattern tree, which prevents them to appear in the result. Generalized Tree Patterns (GTPs) extend classical TAX pattern trees by creating groups of nodes to facilitate their manipulation and by enriching edges to be extracted by the \textit{mandatory/optional} matching option \cite{ChenJLP03}. 

An option more than a limitation of TAX pattern trees  is that a set of similar nodes of the same subtree appears in the resulting tree. For example, a book written by more than one author results from a matching of a pattern containing a single author node. Edges of Annotated Pattern Trees (APTs) \cite{PaparizosWLJ04} solve this problem and present four matching specifications: one to many matches (+), one match only (-), zero to many matches (*) and zero or one match (?).             

APTs, like TAX pattern trees and GTPs, preserve the order of nodes from the input XML data in the output (result), whatever the order of nodes in the pattern tree. To avoid this issue, it is necessary to specify node order in the pattern tree. APTs used in the TLC (Tree Logical Classes) Select and Join operators~\cite{PaparizosWLJ04} are extended with an order parameter (\textit{ord}) \cite{PaparizosJ05}.

We recapitulate in Table~\ref{tab-comparison-modeles-arbres} the characteristics of all pattern trees studied in this section.

\begin{table*}[!t]
\caption{Comparison of pattern trees used in XML tree algebras}
\label{tab-comparison-modeles-arbres}
\begin{center}
\begin{tabular}{|l|c|c|c|c|}  
\hline & \textbf{Matching features} & \textbf{Reordering} & \textbf{Hierarchies}\\
\hline
\textbf{TAX PT} \cite{JagadishLST01} 
& Basic & No & No \\
\hline
\textbf{GTP} \cite{ChenJLP03} 
& Mandatory/optional edges & No & No \\
\hline
\textbf{APT} \cite{PaparizosWLJ04} 
& Edge cardinality & No & No \\
\hline
\textbf{Ordered APT} \cite{PaparizosJ05} 
& Order specification & Yes & No \\
\hline
\end{tabular}
\end{center}
\end{table*}

\section{Complex Hierarchies}
\label{complex-hierarchies}

In this section, we define what we term complex hierarchies. To this aim, we first formalize the definitions of data warehousing concepts.

\subsection{Data Warehouses}
\label{data-warehouses}

\subsubsection{Data Warehouse}
\label{def-data-warehouse}

A data warehouse $W$ modeled w.r.t. a snowflake schema (i.e., with dimension hierarchies) is defined as $W = (F, \cal{D})$, where:
\begin{itemize}
	\item $F$ is a set of facts to observe;
	\item $\cal{D}$ is a set of dimensions or analysis axes. Let $d = |\cal{D}|$.
\end{itemize}

\subsubsection{Dimension and Hierarchy} 
\label{def-dim-and-hierarchy}

$\forall i \in [1, d]$, a dimension $D_i \in \cal{D}$ is defined as a hierarchy made up of a set of $n_i$ levels: $D_i = \{H_{ij}\}_{j = 1, n_i}$. By convention, we denote $H_{i1}$ as the lowest granularity level.

$\forall j \in [1, n_i]$, a hierarchy level $H_{ij}$ is defined in intention as $H_{ij} = (ID_{ij}, \{A_{ijk}\}_{k = 1, a_{ij}}, R_{ij})$, where:
\begin{itemize}
	\item $ID_{ij}$ is the identifier attribute of $H_{ij}$;
	\item $\{A_{ijk}\}$ is a set of $a_{ij}$ so-called member attributes of $H_{ij}$;
	\item $R_{ij}$ is an attribute that references a hierarchy level at a higher granularity than that of $H_{ij}$ (notion of \emph{rollup}).
\end{itemize}

Let $dom()$ be a function that associates to any
attribute its definition domain. Let $h_{ij} =
|H_{ij}|$. $\forall l \in [1, h_{ij}]$, instances
of $H_{ij}$ are tuples under form $(\sigma_{ijl},
\{\alpha_{ijkl}\}_{k = 1, a_{ij}}, \rho_{ijl})$, where:
\begin{itemize}
\item $\sigma_{ijl} \in dom(ID_{ij})$;
\item $\alpha_{ijkl} \in dom(A_{ijk})$ $\forall k \in [1, a_{ij}]$;
\item $\rho_{ijl} \in dom(ID_{ij^{~'}})$ with $j^{~'} \in [1, n_i]$.
\end{itemize}

\subsubsection{Fact}
\label{Def-Fact}

$F$ is defined in intention as $F = (\{\Delta_i\}_{i = 1, d}, \{M_j\}_{j = 1, m})$, where:
\begin{itemize}

	\item $\{\Delta_i\}$ is a set of $d$ attributes that reference instances of hierarchy levels $H_{i1}$ of each dimension $D_i \in \cal{D}$;
	\item $\{M_j\}$ is a set of $m$ measure (or indicator) attributes that characterize facts.

\end{itemize}

Let $f = |F|$. $\forall k \in [1, f]$, instances
of $F$ are tuples under form $(\{\delta_{ik}\}_{i
= 1, d}, \{\mu_{jk}\}_{j = 1, m})$, where:
\begin{itemize}
\item $\delta_{ik} \in dom(ID_{i1})$ $\forall i \in [1, d]$;
\item $\mu_{jk} \in dom(M_j)$ $\forall j \in [1, m]$.
\end{itemize}

\subsection{Complex Hierarchies}
\label{Def-Complex-Hierarchy}

A dimension hierarchy $D_i$ is termed complex if it is both non-strict and non-covering.

\subsubsection{Non-Strict Hierarchy}
\label{sec:HierarchieNonStricte}

A hierarchy is non-strict \cite{Abello06ss,MalinowskiZ06,Tor03} or multiple-arc \cite{Riz07} when attribute $R_{ij}$ is multivalued. In other terms, from a conceptual point of view, a hierarchy is non-strict if the relationship between two hierarchical levels is many-to-many instead of one-to-many. For example, in a dimension describing products, a product may belong to several categories instead of just one.

Similarly, a many-to-many relationship between facts and dimension instances may exist \cite{Riz07}. For instance, in a sale data warehouse, a fact may be related to a combination of promotional offers rather than just one. Formally, here, attributes $\Delta_i$ $(\forall i \in [1, d])$ may be multivalued. 

\subsubsection{Non-Covering Hierarchy}
\label{sec:HierarchieNonCouvrante}

A hierarchy is non-covering \cite{Abello06ss,MalinowskiZ06,Tor03} or ragged \cite{Riz07} if attribute $R_{ij}$ allows linking a hierarchy level $H_{ij}$ to another hierarchy level $H_{ij^{~'}}$ by ``skipping'' one or more intermediary levels, i.e., $\rho_{ij} = \sigma_{ij^{~'}}$ and $\exists H_{ij^{~''}} \in D_i$ / $\rho_{ij^{~''}} = \sigma_{ij^{~'}}$. This occurs, for instance, if in a dimension describing stores, the store-city-region-country hierarchy allows a store to be located in a given region without being related to a city (stores in rural areas). 

Similarly, facts may be described at heterogeneous granularity levels. For example, still in our sale data warehouse, sale volume may be known at the store level in one part of the world (e.g., Europe), but only at a more aggregate level (e.g., country) in other geographical areas. This means that $\forall i \in [1, d]$, $\delta_i \in dom(ID_{ij})$ with $j \in [1, n_i]$ (constraint $j = 1$ is forsaken).

\subsubsection{Notes}
\label{sec:Notes}

\begin{itemize}
	\item The notion of ragged hierarchy has different meanings in the literature. For example, Beyer et al. define it as a hierarchy that is both non-strict and non-covering \cite{BeyerCCOPX05}, while Rizzi defines it as non-covering only \cite{Riz07}. This is why we prefer and define the new terms of complex hierarchy. Malinowski and Zim{\`{a}}nyi use similar switchable terms: generalized hierarchy \cite{MalinowskiZ06} and complex generalized hierarchy \cite{Malinowski08z}. However, these hierarchies include non-covering hierarchies, but not non-strict hierarchies.
	\item Taking complex hierarchies into account involves important summarizability issues \cite{MazonLT09}. However, taking them into account in an XOLAP context is relevant (real cases do exist) and necessary. Research devoted to normalizing conceptual models with summarizability problems \cite{MazonLT08} could be exploited for this sake.
\end{itemize}

\subsection{Example}
\label{sec:example}

\begin{figure*}[!t]
\centering
\includegraphics[width=13cm]{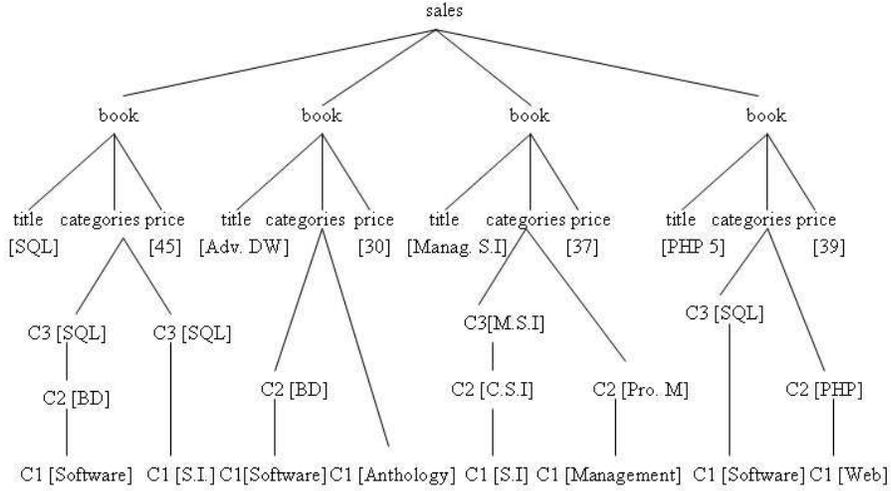}
\caption{Book sales expressed in complex hierarchy}
\label{Sales-Complex-Hierarchies.jpg}
\end{figure*}

Let us expand the example from Figure~\ref{Doc-Books.jpg} with Figure~\ref{Sales-Complex-Hierarchies.jpg}, where book sales are described by titles, categories and sale prices. 

\begin{figure*}[!t]
\centering
\includegraphics[width=10cm]{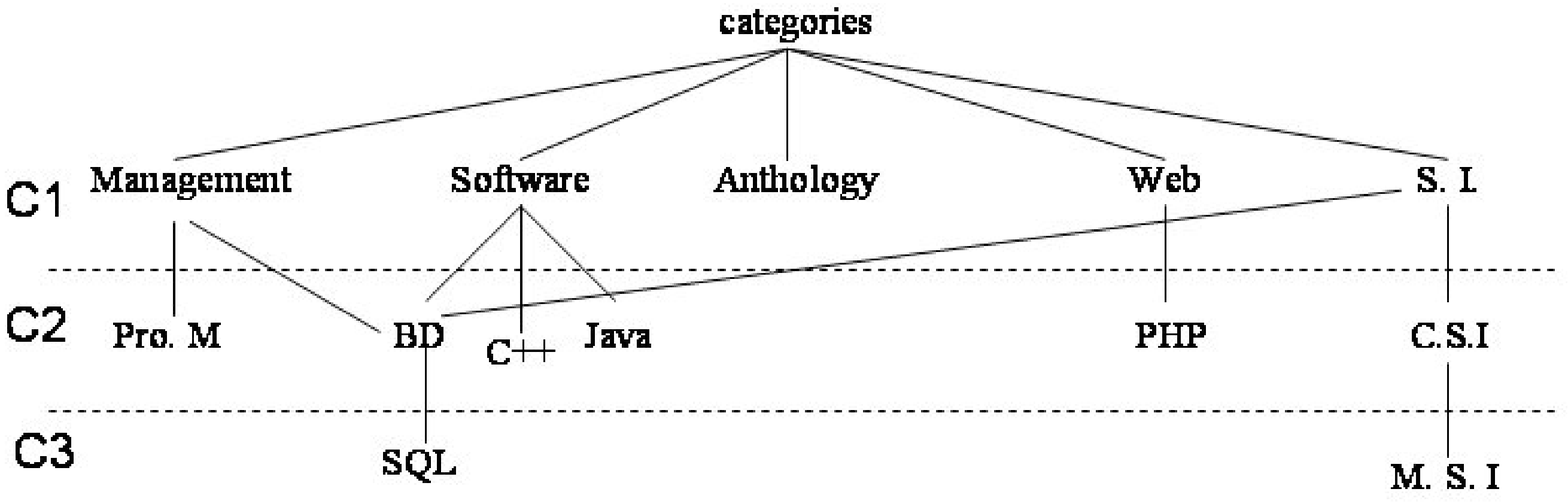}
\caption{Category complex hierarchy}
\label{Complex-Hierarchy.jpg}
\end{figure*}

Each category is associated to a hierarchy level labeled C1 to C3, from the most detailed to the most general. Categories form a complex hierarchy (Figure~\ref{Complex-Hierarchy.jpg}). A category includes more than one book and a book is described by more than one category, thus making this hierarchy non-strict. Moreover, two books (title = ``SQL'' and ``Manag. S.I'') are described by complete hierarchies of categories (C3$/$C2$/$C1). While book entitled ``PHP 5'' is described by an incomplete hierarchy of categories (C3[SQL]$//$C1[Software]). Book entitled ``SQL'' is also described by two hierarchies (one complete and one incomplete). Hence, the hierarchy of categories is non-covering. Being also non-strict, it is thus complex.

\section{Pattern Tree-based Rollup Operator}
\label{A-Pattern-Tree-for-XOLAP}

\subsection{Motivation}
\label{problematic}

In a previous work, we expressed classical OLAP operators with a succession of TAX operators of selection, grouping, join, aggregation and node~update \cite{datax08hmd}. Multidimensional XML data introduced in this work were described by simple (strict with no overlap between levels) hierarchies.

The problem with complex hierarchies is that, when aggregating data, we handle facts described with respect to various levels of granularity. It is then be difficult, in this case, to aggregate measures. A second issue is that some data may not be taken into account because of missing levels in non-covering hierarchies (e.g., book entitled ``PHP 5'').

Choosing to extend the pattern tree of one or more TAX operators used to express the rollup operator (selection, grouping, join, aggregation and node update) from \cite{datax08hmd} is probably a good but not generic solution, since multiple possibilities of extension are possible. For example, we can employ a pattern tree adapted to complex hierarchies in the input of the TAX selection operator, but also join initial data with complex hierarchies using an adapted pattern tree. Furthermore, the TAX pattern tree and its extensions do not take hierarchies into account  (Table~\ref{tab-comparison-modeles-arbres}). Thus, we had to handle them in a separate representation \cite{datax08hmd}.

Since we aim to define a formal framework for XOLAP, rather than extending one or multiple TAX operators, we propose a new rollup operator based on a pattern tree independent from TAX and respecting the definition of rollup \cite{ChaudhuriD97}. 

In the following two sections, we detail our XOLAP rollup operator by presenting the proposed pattern tree and the algorithm allowing to aggregate multidimensional XML data expressed in complex hierarchies using this pattern. This rollup operator inputs a multidimensional XML data tree and outputs a second a multidimensional XML data tree  where measures are aggregated. It is based on an algorithm allowing to match a pattern tree against a multidimensional XML data tree. 

\subsection{Pattern Tree for Rollup}
\label{pattern-tree-of-roll-up}

\begin{figure*}[!t]
\centering
\includegraphics[width=9cm]{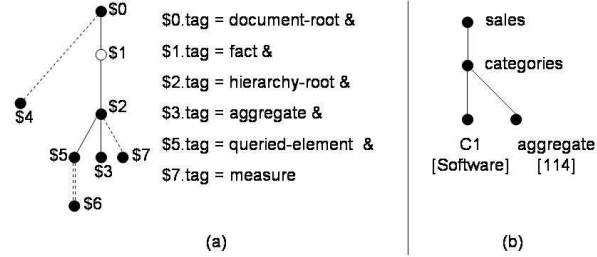}
\caption{Pattern tree for rollup~(a) and witness tree (result)~(b)}
\label{patternforalgo.jpg}
\end{figure*}

We detail here the structure of our pattern tree~(Figure~\ref{patternforalgo.jpg}(a)). 
The graph on the left-hand side represents the pattern tree, while the right-hand side of the figure features formula $F$. 

Parent-child (\textit{pc}) relationships are represented with single edges (/);~ancestor-descendant (\textit{ad}) relationships are represented with double edges (//);~nodes with a white background (\$1, \$4, \$6 and \$7) do not appear in the result, unlike nodes with a black background;~nodes connected to their parent by dotted edges (\$4) are not used in the matching process since they do not have an equivalent in the data tree, unlike nodes connected to their parent nodes by solid edges. 

In Formula $F$,~\$0 is the root of the fact document.~\$1 is a fact described by its dimensions and measures.~\$2 is the root of the complex hierarchy.~\$3 computes aggregation from measure~\$7 of each fact~\$1.~\$4 counts the number of matched facts. It is useful for aggregation operations such as average.~\$5~is the most detailed element of the hierarchy, child (direct descendant) of~\$2 in the data tree.~\$6 is any descendant of~\$5.~\$6 is used to browse through the hierarchy in the matched data tree.

\subsection{Rollup Algorithm}
\label{Algo-detailed}

Our rollup algorithm (Algoritm~\ref{algo-roll-up}) is based on the pattern tree from Figure~\ref{patternforalgo.jpg}(a). For each fact~\$1, it checks whether the highest hierarchical element (direct child \$5 of the hierarchy root \$2) corresponds to the input hierarchical element \emph{H-el-agg}. If so, aggregation is computed from measure~\$7, \$3 and \$4 are updated (they are input-output parameters of function \emph{AGGREGATE}), and the algorithm steps to the next fact. Otherwise, the algorithm continues to scan through the hierarchy \$6 until finding the hierarchical element to be aggregated. 
 
\begin{algorithm}
\label{algo-roll-up}
\caption{Rollup}
\begin{algorithmic}
\begin{small}
\STATE \textbf{Input:} DT // Fact data tree
\STATE ~~~~~~~~~H-el-agg // Hierarchical element to be aggregated
\STATE \$3~$\leftarrow$~0 
\STATE \$4~$\leftarrow$~0 
\STATE Stop~$\leftarrow$~FALSE
\FORALL {\$1 in DT} 
  \WHILE{exists child~\$5~of~\$2 \textbf{and not} Stop}  
     \IF{\$5.value = H-el-agg}        
         \STATE AGGREGATE (\$3, \$4, \$7)              
         \STATE Stop~$\leftarrow$~TRUE  
         \ELSIF{\$6.value = H-el-agg} 
         \STATE AGGREGATE (\$3, \$4, \$7)              
         \STATE Stop~$\leftarrow$~TRUE     
      \ENDIF
    \ENDWHILE
\ENDFOR
\end{small}
\end{algorithmic}
\end{algorithm}

\subsection{Example}
\label{Exp-roll-up-based-on-Algo}

Let us consider query~Q:~``compute total of book sales for category Software''. Category Software means C1[Software] or any descendant category. The aggregate function used in \emph{AGGREGATE} (Algorithm~\ref{algo-roll-up}) in this case is $sum$. When matching the pattern tree from Figure~\ref{patternforalgo.jpg}(a) against the data tree from Figure~\ref{Sales-Complex-Hierarchies.jpg}, pattern node \$0 takes the value ``sales'' and~\$1 takes the value ``book''.~\$2 must be equal to~``categories'' here. For each \$2 = ``categories'' of every ``book'', we check whether the value of~\$5 (the most detailed category of the book) is equal to the looked for hierarchical value (\emph{H-el-agg} = Software). If it is true, we step to the next book (next \$1). Otherwise, we continue to check whether one descendant of~\$5~(\$6, an ancestor category) corresponds to category Software. In case a book of category ``Software'' is found, \$4 is incremented and \$3 is incremented by measure value \$7 of this book (to compute sale total). When Software is not found after searching for all categories of the current~\$1, we step to the next fact (\$1 = book) and continue searching. After matching all the data tree with the pattern tree, the aggregate is computed,~\$5 takes the value of the searched category Software and~\$3 the computed total book sale, Aggregate, as shown in Figure~\ref{patternforalgo.jpg}(b), which represents the witness tree answering our initial query.

\section{Conclusion and Perspectives}
\label{Conclusion}

In this paper, we propose to the best of our knowledge the first pattern tree for multidimensional data since the introduction of pattern trees in XML approaches \cite{AmerYahiaCLS01}. Though it is simple, this pattern tree permits to aggregate data expressed in complex hierarchies, no matter their structure. We thus progressed toward the definition of a formal framework for XOLAP. It is important that XML multidimensional data are processed natively, which allows taking into account XML specifics such as complex hierarchies, which are intricate to handle in relational systems.

The perspectives of this work are twofold. We aim, in a first step, to adapt the principle of the pattern tree we introduce in this paper to other XOLAP operators (cube, drill down, etc.) in order to complete our algebra. More matching options (e.g., optional edges or edge cardinalities) might have to be added to the pattern tree model at this stage. Moreover, XOLAP operators performing aggregation raise summarizability problems. We aim to present solutions to detect and correct them in the algorithms (and patterns) associated to the different operators. 

In a second stage, we plan to implement our algebra (as a proof of concept) and optimize its performance. Pattern tree-based XQuery optimization approaches may help optimize our operators under their physical form. For instance, we could use minimization techniques. Minimizing a pattern tree $pt$ consists in constructing a minimal pattern that is equivalent to $pt$ while bearing the minimum possible size \cite{AmerYahiaCLS01}.

\bibliographystyle{IEEEtran}
\bibliography{icmwi10-hachicha-darmont}

\end{document}